\documentclass[rmp,aps,showkeys]{revtex4-1}

\usepackage{multirow}
\usepackage{array}
\usepackage{amsmath}
\usepackage{graphicx}% Include figure files
\usepackage{dcolumn}% Align table columns on decimal point
\usepackage{bm}% bold math
\usepackage{verbatim}
\DeclareMathAlphabet \mathbfcal{OMS}{cmsy}{b}{n}
\begin{document}

\title{Interacting Dirac fermions and the rise of Pfaffians in graphene}

\author{Vadym Apalkov$^1$ and Tapash Charkaborty$^{2,3}$}
\affiliation{$^1$Department of Physics and Astronomy, Georgia State
University, Atlanta, Georgia 30303, USA;\\ $^2$Department of Physics and Astronomy,
University of Manitoba, Winnipeg, Canada R3T 2N2;\\
$^3$Department of Physics, Brock University, St. Catharines, ON, Canada L2S 3A1}

\date{\today}

\begin{abstract}

Fractional Quantum Hall effect (FQHE) is a unique many-body phenomenon, which was discovered 
in a two-dimensional electron system placed in a strong perpendicular magnetic field. It is 
entirely due to the electron-electron interactions within a given Landau level. For special 
filling factors of the Landau level, a many-particle incompressible state with a finite 
collective gap is formed. Among these states, when the Landau level is half filled, there 
is a special FQHE state that is described by the Pfaffian function and the state supports 
charged excitations that obey non-Abelian statistics. Such a $1/2$-FQHE state can be realized 
only for a special profile of the electron-electron potential. For example, for conventional 
electron systems the $1/2$-FQHE state occurs only in the second Landau level, while in a 
graphene monolayer, no $1/2$-FQHE state can be found in the first and the second Landau 
levels. Another type of low-dimensional system is the bilayer graphene, which consists of 
two graphene monolayers coupled through the inter-layer hopping. The system is quasi-two-dimensional, 
which makes it possible to tune the inter-electron interaction potential by applying either 
the bias voltage or the magnetic field that is applied parallel to the bilayer. 
Interestingly, in the bilayer graphene with AB staking, there is one Landau level per 
valley where the $1/2$-FQHE state can indeed be present. The properties of that $1/2$-FQHE state 
have a nonmonotonic dependence on the applied magnetic field and this state can be even more stable 
than the one discovered in conventional electron systems. 
\end{abstract}
%\pacs{}
\maketitle

\section*{Key points}
\begin{itemize}
\item In two-dimensional electron systems an externally applied magnetic field results in 
the formation of highly discrete and degenerate Landau levels.
\item In a strong enough magnetic field, the discrete nature of the energy spectrum of 
two-dimensional electrons results in the Integer Quantum Hall effect, which happens when integer 
number of Landau levels are occupied.
\item For a partially filled Landau levels, inter-electron interactions result in formation 
of incompressible liquids, which manifest themselves as the Fractional Quantum Hall Effect.
\item The conventional FQHE corresponds to filling factors of type $p/q$, where $q$ is odd.  
\item The unconventional FQHE at half-filling ($\nu = 1/2$) of a given Landau level is 
described by the Pfaffian function as the wave function of the ground state and has excitations 
with the charge of $e/4$ and obey non-abelian statistics. 
\item For conventional 2D systems, $1/2$-FQHE is realized only in the $n=1$ Landau level. 
\item For graphene monolayer, there is no $\nu =1/2$ incompressible liquid in any Landau level.
\item For bilayer graphene there is a special Landau level in which the stability of the $\nu =1/2$ 
incompressible state can be tuned by the magnitude and direction of the magnetic field and the bias voltage.   
\end{itemize}

\section{Introduction}

A moving charge placed in an external magnetic field experiences a magnetic force, which 
can change only the direction of the charge velocity but not its magnitude. One of the 
manifestations of this property of the magnetic field is the Hall effect. The Hall effect 
occurs when the current flows through a solid and the magnetic field is applied perpendicular 
to the current. In this case the voltage difference, which is called the Hall voltage, 
is generated across the conductor in the direction transverse to both the current and 
the magnetic field. This effect was discovered by Edwin Hall in 1879 \cite{Hall_1879}. 
If the electric current is in the $x$ direction with the current density of $j^{}_x$ and 
magnetic field is in  the $z$ direction, then there is a Hall electric field generated 
in the $y$ direction. The strength of the Hall effect is characterized by the Hall 
coefficient defined by the following expression \cite{Solid_Ashcroft,Hurds_1972,Solid_Kittel} 
\begin{equation}
R^{}_H = \frac{E^{}_y}{j^{}_x B^{}_z}.
\label{Hall0}
\end{equation}
It is possible to show that, within the classical approach, the Hall constant is related to 
the charge of the carriers, $q$, and their density, $n$, 
\begin{equation}
R^{}_H = \frac{1}{nq}.
\label{Hall1}
\end{equation}
One important property of this expression is that the Hall constant and correspondingly the 
Hall voltage depends on the sign of carrier's charge, $q$. For example, for semiconductors, 
when the carriers are negatively charged electrons or positively charged holes, the Hall 
constant can be  negative or positive depending on the type of the major carriers.  

The above expression for the Hall coefficient corresponds to the classical description of 
the electron dynamics in a magnetic field, or in the low magnetic field limit. With increasing 
magnetic field, the electron dynamics in the magnetic field becomes essentially quantum mechanical, 
which results in quantization of the corresponding energy spectrum and formation of highly 
degenerate and discrete Landau levels \cite{Landau_Lifshitz_Quantum_Mechanics_1965}. The 
energy distance between the Landau levels is proportional to the magnetic field. Due to 
discrete nature of the energy spectrum of carriers, the diagonal resistance of a solid 
as a function of the magnetic field shows oscillations, which are called the Shubnikov de Haas 
oscillations \cite{Shubnikov_de_Haas}. 

More interesting and unexpected phenomena occur in two-dimensional (2D) electron systems, such 
as in heterojuctions, atomic monolayers, or in quantum wells. If the magnetic field is applied 
perpendicular to the 2D layer, i.e., the $(x,y)$ plane, then for weak magnetic fields there is 
the classical Hall effect, which can be also characterized by the Hall resistance $R^{}_{xy}$
\begin{equation}
R^{}_{xy} = \frac{E^{}_y}{j^{}_x }= R^{}_{H}B^{}_z.
\label{Hall2D}
\end{equation}
Therefore in the classical regime, i.e., in a weak magnetic field, the Hall resistance is proportional 
to the magnetic field, $B^{}_z$. At the same time, in stronger magnetic fields when the quantization 
of the in-plane electron dynamics results in Landau levels, there is another unique effect, which 
is the Integer Quantum Hall Effect (IQHE) \cite{Stone_QHE_book,QHE_PRL_1980,Yoshioka_QHE_book_2002,%
Prange_Girvin_book_QHE_1987,Chakraborty_book_QHE_1995,QHE_review_2017,QHE_40_year_2020}. In this regime, 
the Hall resistance takes only quantized values of the form
\begin{equation}
R_{xy} = \frac{h}{e^2 \nu },
\label{Rxy}
\end{equation}
where $h$ is the Plank constant, $e$ is the electron charge, and $\nu $ is an integer, $\nu = 1,2,3, 
\ldots$. The integer number $\nu$ corresponds to the number of completely filled Landau levels and 
whenever the Fermi energy is between the two Landau levels $n$ and $n+1$, the Hall resistance is 
constant, while the longitudinal resistance, $R^{}_{xx}$, becomes zero. Strictly speaking, to 
understand the Quantum Hall Effect we also need to consider broadening of the Landau levels due 
to disorder and spatial localization of the in-gap electron states between the Landau levels 
\cite{Stone_QHE_book,Chakraborty_book_QHE_1995}. However the main condition for the IQHE is the 
existence of the gaps in the energy spectrum which naturally occurs in 2D systems in a strong magnetic field. 

 \begin{figure}
\begin{center}\includegraphics[width=0.47\textwidth]{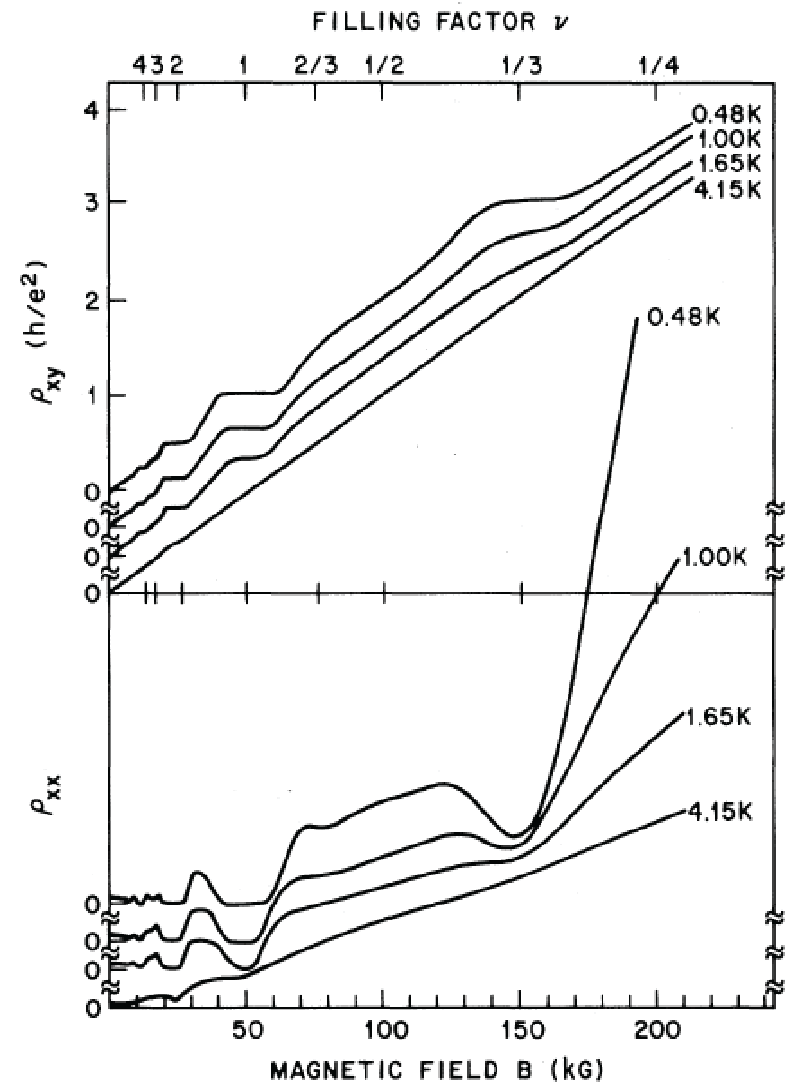}\end{center}
%\vspace*{-1cm}
\caption{ Non-diagonal and diagonal resistivities, $\rho^{}_{xy}$ and $\rho^{}_{xx}$, as a function of 
magnetic field, $B$. The sample is GaAs-Al$^{}_{0.3}$Ga$^{}_0.7$As with electron density of 
$n= 1.23\times 10^{11}$ cm$^{-2}$  and mobility $\mu = 90000$ 
cm$^2$/Vs (Adapted from \cite{FQHE_first_observation}). }
\label{FQHE_first}
\end{figure}

The IQHE was discovered experimentally in 1980 and two years later another unexpected (and more
spectacular) effect was observed in even stronger magnetic fields, and in better quality samples. 
This effect is known as the Fractional Quantum Hall Effect (FQHE) \cite{FQHE_first_observation,%
Chakraborty_book_QHE_1995}. In this case the filling factor $\nu $ in Eq.~(\ref{Rxy}) is fractional  
and corresponds to partial occupation of a given Landau level. Since all levels within the Landau 
level are degenerate, the electron-electron interaction completely determines the properties of the 
system. At special fractional occupations of the Landau level, the interaction generates the incompressible 
ground states with gapped excitations, which finally results in the FQHE. The primary filling factors 
where the FQHE occur are $\nu =\frac{1}{3}, \frac{2}{3},\frac{1}{5},\frac{2}{5}, \frac{3}{5}, 
\frac{1}{7}$ etc. \cite{Prange_Girvin_book_QHE_1987,Chakraborty_book_QHE_1995,Halperin_Jain_book_FQHE}. 
In all these cases the denominator is an odd number, which is due to the fermionic nature of electrons.  
  
\section{Fractional Quantum Hall Effect}

First, we consider the conventional two-dimensional electron systems, such as the ones grown in heterojuncions 
or in a quantum well, for which the low-energy dispersion relation is parabolic. The Hamiltonian of such 
systems has the following form 
\begin{equation}
{\cal H}^{}_c = \frac{\vec{p}^2}{2m^*},
\label{Hc1}
\end{equation}
where $\vec{p}$ is the two-dimensional momentum and $m^*$ is the electron effective mass. When the 
system is placed in a perpendicular external magnetic field, the Hamiltonian becomes 
\begin{equation}
{\cal H}^{}_c = \frac{\mathbf{\pi }^2}{2m^*},
\label{Hc2}
\end{equation}
where $\vec{\pi} = \vec{p} + e \vec{A}/c$ is the generalized momentum and $\vec{A}$ is the vector 
potential. The energy spectrum of the Hamiltonian (\ref{Hc2}) consists of highly degenerate discrete 
Landau levels \cite{Landau_Lifshitz_Quantum_Mechanics_1965} with energy 
\begin{equation}
\epsilon^{}_n = \hbar \omega^{}_B \left( n + \frac{1}{2} \right),
\label{LConv}
\end{equation}
where $\omega^{}_B = e B/m^*$ is the cyclotron frequency and $n = 0,1,2,\ldots$ is the Landau level 
index. We write the corresponding Landau eigenfunctions as $\phi^{}_{n,m}$, where the index 
$m$ labels the degenerate wavefunctions within a given Landau level and is determined by the choice 
of the gauge. For example, in the Landau gauge, $A^{}_x =0$ and $A^{}_y= Bx$, the index $m$ is the 
$y$-component of the momentum, while in the symmetric gauge, $\vec{A} = \frac12\vec{B}\times \vec{r}$, 
the index $m$ is the $z$-component of electron angular momentum \cite{Landau_Lifshitz_Quantum_Mechanics_1965}.
In what follows, we consider $m$ as the $z$ component of the angular momentum.  
 
When a given Landau level is partially occupied by electrons, the properties of the system is completely 
determined by the electron-electron interactions. If the mixing of Landau levels due to the interactions 
is weak, then the interaction properties of electrons within a single Landau level  are completely 
determined by the Haldane pseudopotentials $V^{(n)}_m$ \cite{Haldane_PRL_1983}. The Haldane pseudopotential  
$V^{(n)}_m$ is the energy of two electrons with the relative angular momentum $m$ and they can be  
found from the following expression \cite{Haldane_in_book} 
\begin{equation}
V_m^{(n)} = \int_0^{\infty } \frac{dq}{2\pi} q V(q)
\left[F^{}_n(q) \right]^2 L^{}_m (q^2)
 {\rm e}^{-q^2},
\label{Vm}
\end{equation}
where $L^{}_m(x)$ are the Laguerre polynomials, $V(q) = 2\pi {\rm e}^2/(\kappa \ell^{}_0 q)$
is the Coulomb interaction in the momentum space, $\kappa$ is the dielectric constant, 
and $F^{}_n(q)$ is the form factor of the $n$-th Landau level. The form factor is totally determined 
by the structure of the wavefunctions of the corresponding Landau level. Different two-dimensional 
systems, such as graphene monolayer, transition-metal dichalcogenide monolayer, graphene bilayer, etc.,
have different form factors $F^{}_n(q)$, but in all cases, the corresponding Haldane pseudopotentials 
are given by the same expression (\ref{Vm}). For conventional two-dimensional electron systems with Landau 
wavefunctions $\phi^{}_{n,m}$ the form factor is \cite{Haldane_in_book}
\begin{equation}
F^{}_n(q) =  L^{}_n\left( q^2/2\right).
\label{FF1}
\end{equation}

If the electron system is fully spin polarized then the spatial component of the many-particle 
wavefunction is antisymmetric with respect to the particle exchange. In this case, only the Haldane 
pseudopotentials, $V_m^{(n)}$, with odd values of $m$, $m=1,3,5,\ldots$, determine the properties 
of the system. The nature of the ground state mainly depends on the ratio of the first few 
pseudopotentials, $V_1^{(n)}/V_3^{(n)}$ and $V_3^{(n)}/V_5^{(n)}$.

%Below all pseudopotentials are given in units of the Coulomb energy, $\varepsilon^{}_C = e^2 /\kappa\ell^{}_0$.

At special values of the filling factor $\nu$, which is defined as $\nu = N^{}_e/N^{}_0$, where 
$N^{}_e$ is the number of electrons in a given Landau level and $N^{}_0$ is its degeneracy, the 
electron system becomes incompressible with a finite excitation gap. This incompressibility is 
entirely due to electron-electron interactions and is determined by the values of the Haldane 
pseudopotentials. Only for these filling factors the FQHE can be observed. 
 
The conventional FQHE occurs at the filling factors of the form $p/q$, where $q$ is odd, which is 
related to the antisymmetric property of a many-particle electron wavefunction when two particles 
are interchanged, i.e., $\Psi (\vec{r}^{}_1,\vec{r}^{}_2, \vec{r}^{}_3,\ldots) = - \Psi 
(\vec{r}^{}_2,\vec{r}^{}_1, \vec{r}^{}_3,\ldots)$. A prominent example of this type of FQHE is the
$1/q$ sequence, i.e., $1/3$, $1/5$ etc. The corresponding many-particle wave function is well described 
by the Laughlin  function \cite{Laughlin_PRL_1983} of the form 
\begin{equation}
\Psi(z^{}_1, z^{}_2, z^{}_3,\ldots) = \prod^{}_{i<j} (z^{}_i - z^{}_j)^q 
\exp\left(-\sum^{}_i\frac{z_i^2}{4\ell_0^2}\right),
\label{LF}
\end{equation}
where the positions of electrons are described in terms of complex variable $z = x-iy$. The 
Laughlin function (\ref{LF}) corresponds to the filling factor $1/q$ and is very close to the exact 
many-particle wave function of the system. The accuracy of the trial wave functions, for instance
the Laughlin functions is tested through numerical calculations for a finite-size electron system. 
For a finite-size system placed in an external magnetic field, the many-particle basis is finite, 
which allows us to find both the ground state and the excitation spectrum of the system.  Then, the 
overlap of the trial function with the exact ground state wave function of the many-particle system 
can be determined and that can be used to test the accuracy of the trial approximation. The 
incompressibility of the system is determined by the  magnitude of the gap in the collective excitations 
of the system. Another interesting property of the $1/q$ FQHE, which is described by the Laughlin 
function, is that the charged excitations have fractional charge $e^*=e/q$. For example, for the $1/3$-FQHE
the quasiparticles carry a fractional charge $e^*=e/3$.

Our fundamental understanding of the origin of the odd-denominator FQHE is from the brilliant work 
of Laughlin which requires that we consider a {\it incompressible} fluid that has no long-range
positional order. The incompressibility implies that all the excited states have a non-zero energy
difference from the ground state. The ground state can be fully spin polarized or also have spin degree
of freedom \cite{Chakraborty_1984,Chakraborty_PRL_1985,Apalkov_PRL_2001} in the
ground state and spin-reversed excitations. It is worth emphasizing that despite the torrent of ideas
unleased by the Laughlin function, the origin of incompressibility in the Laughlin state
still remains unresolved \cite{Haldane_2011}.

In addition to the conventional FQHE set of type $p/q$, where $q$ is odd, another type of FQHE corresponding 
to the filling $5/2$ has been discovered experimentally \citep{FQHE_5_2_PRL_1987,QHE_5_2_review}. This 
filling factor implies that there are two completely occupied Landau levels, which correspond to spin-up 
and spin-down Landau levels with index $n=0$, and the next Landau level is half-filled. It means 
that in the $n=1$ Landau level the $1/2$-FQHE occurs, i.e., the filling factor of the $n=1$ Landau level 
is $\nu = 1/2$. This unusual FQHE cannot be described by the Laughlin function (\ref{LF}). This is 
because the many-particle wave function should be antisymmetric with respect to the particle interchange,
because the corresponding electrons are fermions and $q$ must be an odd integer. But for $q=2$, 
the Laughlin function is symmetric, and therefore describes a system of bosons. Another 
important property of the $1/2$-FQHE is that the $1/2$ incompressible liquid is realized only 
in higher Landau levels, for example, for $n=1$ but not for the $n=0$ level. Different wavefunctions 
have been proposed theoretically to describe this state. Among them are the Pfaffian \cite{Read_pfaffian}, 
anti-Pfaffian \cite{anti_pfaffian_PRL_2007}, particle-hole symmetric Pfaffian \cite{PH_Pfaffian}, and 
the 221-parton \cite{Non_Abelian_statistics_fractional_quantum_Hall_states} wavefunctions. The $1/2$-FQHE 
is also sensitive to the inter-Landau mixing \cite{LL_mixing_5_2_PRL} which modifies the electron-electron 
interaction potential \cite{realistic_Hamiltonian_5_2} and opens up the possibility to observe the $1/2$-FQHE 
in higher Landau levels \cite{graphene_1_2_high_Landau_exp_2018} where the inter-Landau mixing becomes 
strong. Below we consider the properties of the $1/2$-FQHE, which is described only by the Pfaffian wavefunction. 

\subsection{Pfaffian function}

Some properties of the FQHE states can be understood by looking at the specially created composite objects, 
that are called the composite fermions. The composite fermion is an electron with an even number of flux 
quanta attached to it. These objects still have the same Fermi statistics as original electrons, but 
they feel a different magnetic field. To understand their properties, first, let us consider a fully filled
Landau level. In this case, there is one magnetic flux quantum per each electron and under this condition 
we have the IQHE. Now let us consider a Landau level with the filling factor of $\nu =1/3$, i.e., only 
$1/3$ states within a given Landau level are occupied. For such a system, there are three magnetic flux 
quanta per electron and this state corresponds to the $\nu =1/3$-FQHE. The composite fermions for this
system are electrons bound to two magnetic flux quanta each. Therefore, for these composite fermions 
(one electron plus two flux quanta), there is one magnetic flux quanta per fermion, which corresponds 
to a fully occupied Landau level and is therefore, the IQHE of composite fermions. Another way to state
this is that, the FQHE for electrons is the IQHE for composite fermions. This correspondence exists only 
for the FQHE of type of $p/q$, where $q$ is odd. As an example, for $\nu = 1/5$ the composite fermion is 
an electron plus four magnetic flux quanta and the composite fermions all occupy the first Landau level.
It is important that the number of magnetic flux quanta attached to an electron is even. Only in this 
case the composite particles have the same Fermi statistics as electrons. 

A unique situation occurs for the Landau level that is exactly half filled. For such a system, the composite
fermion picture dictates that there are two magnetic flux quanta associated with each electron. In that 
case, a composite fermion comprise of an electron bound to two magnetic flux quanta and in a mean-field
picture, the system of composite fermions does not experience a net magnetic field \cite{Halperin_Lee_Read_PRB}. 
The properties of this composite fermion system are determined by the inter-fermion interaction, i.e., by 
the Haldane pseudopotentials. For the $n=0$ Landau level, the composite fermions forms a gapless composite 
Fermi liquid. A more interesting situation happens for the $n=1$ Landau level, i.e., for the total filling 
factor $\nu =5/2$. In this case, the theoretical and experimental results suggest that the system forms 
a gapped state which is due to pairing of the composite fermions, i.e., by formation of Cooper pairs, 
similar to what we see in the supercoductor systems. The pairing symmetry can then be different and below 
we consider only one type of symmetry, i.e., when the Cooper pairs of composite fermions have angular momentum 
$-1$. It was proposed that the corresponding ground state of the system expressed in terms of the coordinates 
of original electrons is described by the Pfaffian \cite{Read_pfaffian,PRL_5_2_numerical_Greiter,%
Paired_Hall_states_Nuclear_Physics,RMP_nonabelions_FQHE} function, which has the following form
\begin{equation}
\Psi^{}_{\mbox{Pf}}=\mbox{Pf}\left[\frac1{z^{}_i-z^{}_j}\right]         
\prod^{}_{i<j} (z^{}_i-z^{}_j)^2 \exp\left(-\sum^{}_i\frac{z_i^2}{4\ell_0^2}\right),
\label{Pf_12}
\end{equation}
where the Pfaffian is defined as \cite{Read_pfaffian,Landau_level_sphere_Greiter}
\begin{equation}
{\rm Pf}\, \left[ \mathbf{M} \right] =\frac1{2^{N/2}\left(N/2\right)!}\sum^{}_{\sigma
\in S^{}_N}{\rm sgn}\,\sigma\prod_{l=1}^{N/2}M^{}_{\sigma(2l-1)\sigma(2l)},
\end{equation}
for an $N\times N$ antisymmetric matrix $\mathbf{M}$ whose elements are $M^{}_{ij}$, where $N$ is an 
even number. Here $S^{}_N$ is the group of permutations of $N$ objects. 

Historically, the concept of Pfaffians arose from a very surprising mathematical result: Consider the 
determinant of a $N\times N$ antisymmetric matrix $\mathbf{M}=[M^{}_{ij}]$ written as 
$\det [M^{}_{ij}]_{1\leq i,j \leq N}$ in which $M^{}_{ij}= - M^{}_{ji}$ for $i,j = 1, 2, \ldots, N$. 
If $\mathbf{M}= -\mathbf{M}^T$ is an antisymmetric matrix of order $N$ then $\det [\mathbf{M}]= (-1)^N 
\det [ \mathbf{M}]$ so that if the order is odd, then the determinant vanishes. However, when $N= 2m$, 
then the determinant is the square of a polynomial in coefficients of the given determinant. This result 
was first discovered by the British mathematician Arthur Cayley in 1847 \cite{Cayley_1847,Halton_1966}, who named 
this polynomial {\it Pfaffian} in honor of the great German mathematician Johann Pfaff, who discovered 
it in 1815, in the course of his work on differential equations. 

For $N=2$, 
\begin{equation}
\det \left|
\begin{array}{cc}
 0       & M^{}_{12}   \\
 -M^{}_{12} & 0   
\end{array}
\right| = \left( M^{}_{12} \right)^2, 
\end{equation}
and the Pfaffian is ${\rm Pf}\left[ \mathbf{M} \right] = M^{}_{12}$. Similarly, for $N=4$, 
\begin{equation}
\det \left|
\begin{array}{cccc}
 0       & M^{}_{12}  & M^{}_{13} & M^{}_{14} \\
 -M^{}_{12} & 0       & M^{}_{23} & M^{}_{24} \\
 -M^{}_{13} & -M^{}_{23}  & 0      & M^{}_{34}  \\
 -M^{}_{14} & -M^{}_{24}  & -M^{}_{34} & 0     
\end{array}
\right| = \left( M^{}_{12}M^{}_{34} - M^{}_{13}M^{}_{24} +
                     M^{}_{14}M^{}_{23} \right)^2, 
\end{equation}
and the corresponding Pfaffian is 
\begin{equation}
{\rm Pf}\, \left[ \mathbf{M} \right] = 
M^{}_{12}M^{}_{34} -  M^{}_{13}M^{}_{24} +M^{}_{14}M^{}_{23}.
\end{equation}
There are three terms in the above expression. For a general antisymmetric matrix $N\times N$, the 
number of terms in the  pfaffian is $(N-1)!!$. The Pfaffians are sometimes referred to as triangular 
determinant or half-determinant \cite{Half_determinants_1982},
\begin{equation}
{\rm Pf} \, \left[ \mathbf{M} \right] = 
\left. 
\begin{array}{cccc}
 | M^{}_{12}  & M^{}_{13} & M^{}_{14} \\
           & M^{}_{23} & M^{}_{24} \\
           &        & M^{}_{34}      
\end{array}
\right|.
\end{equation}
Moving on, one can obtain the following relation between the determinant of the matrix 
$\mathbf{M}$ and its Pfaffian \cite{Cayley_1847,Halton_1966}, 
\begin{equation}
\rm{det} \left[ \mathbf{M} \right] = \left( {\rm Pf}\, \left[ \mathbf{M} \right] \right)^2.
\end{equation}
This relation can be generalized for a special partially antisymmetric matrix of the form 
\cite{Cayley_1847,pfaffian_properties_PRB_2008}
\begin{equation}
\mathrm{det}
\left[
\begin{array}{ccccc}
 0       & b^{}_{12}  & b^{}_{13} & \ldots & b^{}_{1N} \\
 -a^{}_{12} & 0       & a^{}_{23} & \ldots & a^{}_{2N} \\ 
 -a^{}_{13} & -a^{}_{23} & 0      & \ldots & a^{}_{3N} \\ 
  \vdots & \vdots  & \vdots & \vdots & \vdots \\
 -a^{}_{1N} & -a^{}_{2N}  & -a^{}_{3N} & \ldots & 0 
\end{array}
\right] 
= \mathrm{Pf} 
\left[
\begin{array}{ccccc}
 0       & a^{}_{12}  & a^{}_{13} & \ldots & a^{}_{1N} \\
 -a^{}_{12} & 0       & a^{}_{23} & \ldots & a^{}_{2N} \\ 
 -a^{}_{13} & -a^{}_{23} & 0      & \ldots & a^{}_{3N} \\ 
  \vdots & \vdots  & \vdots & \vdots & \vdots \\
 -a^{}_{1N} & -a^{}_{2N}  & -a^{}_{3N} & \ldots & 0 
\end{array}
\right] 
\times \mathrm{Pf} 
\left[
\begin{array}{ccccc}
 0       & b^{}_{12}  & b^{}_{13} & \ldots & b^{}_{1N} \\
 -b^{}_{12} & 0       & a^{}_{23} & \ldots & a^{}_{2N} \\ 
 -b^{}_{13} & -a^{}_{23} & 0      & \ldots & a^{}_{3N} \\ 
  \vdots & \vdots  & \vdots & \vdots & \vdots \\
 -b^{}_{1N} & -a^{}_{2N}  & -a^{}_{3N} & \ldots & 0 
\end{array}
\right]
\end{equation}

Other useful relations, involving pfaffians of $N\times N$ matrices $\mathbf{M}$ 
and $\mathbf{A}$, are the following 
\begin{eqnarray}
& & {\rm Pf}\, \left[ \mathbf{M}^{T} \right] = (-1)^{N/2}  {\rm Pf}\, \left[ \mathbf{M} \right] \\
& & 
{\rm Pf}\, \left[ \lambda \mathbf{M} \right] = \lambda ^{N/2} {\rm Pf}\, \left[ \mathbf{M} \right] \\
& & {\rm Pf}\, \left[ \mathbf{A} 
\mathbf{M} \mathbf{A}^{T} \right] = 
 {\rm det} \left[ \mathbf{A} \right] {\rm Pf}\, \left[ \mathbf{M} \right].   \\
 & & {\rm Pf} \left[ 
 \begin{array}{cc}
 0 & \mathbf{M} \\
  -\mathbf{M}^{\mathrm{T}} & 0 
 \end{array}
 \right]   = (-1)^{N(N-1)/2} \mathrm{det} \left[ \mathbf{M} \right]   \\
& & 
{\rm Pf} \left[ 
 \begin{array}{cc}
 \mathbf{M}^{}_1 & 0 \\
 0 &  \mathbf{M}^{}_2 
 \end{array}
 \right]   =  \mathrm{Pf} \left[ \mathbf{M}^{}_1 \right] 
 \mathrm{Pf} \left[ \mathbf{M}^{}_2 \right]
\end{eqnarray}

There is also a special expression for the pfaffian of skew-symmetric tridiagonal matrix,
\begin{equation}
\mathrm{Pf} 
\left[
\begin{array}{ccccccc}
 0      & a^{}_{1}  & 0     & 0      &  &        &   \\
 -a^{}_{1} & 0      & 0     & 0      &        &  &  \\  
 0      & 0      & 0     & a^{}_2    &  & &  \\ 
  0     & 0      & -a^{}_2  & 0      &  &&  \\
      &      &       &  & \ddots &  &  \\
       &       &   &       &  & 0 & a^{}_{N/2} \\
      &      &       &  &  & -a^{}_{N/2} &  0      
\end{array}
\right] = a^{}_{1} a^{}_{2} \ldots a^{}_{N/2}
\end{equation}
Pfaffians actually describe a {\it pairing} state. In fact, the famous Bardeen-Cooper-Schrieffer
(BCS) wave function that is the wavefunction for spin-singlet pairs can be expressed in terms of 
the Pfaffians or a determinant \cite{pfaffian_properties_PRB_2008,pairing_wave_functions_determinants}. 

The Pfaffian function (\ref{Pf_12}) which determines the $\nu=1/2$ state is defined only for 
{\it even} number of particles, which illustrates its fundamental property as a collective state of 
the Cooper pairs, i.e., the bound state of two electrons. Interestingly, while the pfaffian itself, 
i.e., the first factor in Eq.~(\ref{Pf_12}), is singular when two electrons are at the same position 
($z^{}_i = z^{}_j$), when multiplied by the second factor, the pfaffian function (\ref{Pf_12}) 
becomes regular. In other words, the Pfaffian state has a nonzero amplitude at the coincidence of
two particles. As an example, for four electrons the pfaffian function takes the following form 
\begin{eqnarray}
& & \Psi^{}_{\mbox{Pf}}=
\left[   
\frac{1}{z^{}_1-z^{}_2} \frac{1}{z^{}_3-z^{}_4} -  
\frac{1}{z^{}_1-z^{}_3} \frac{1}{z^{}_2-z^{}_4} +
\frac{1}{z^{}_1-z^{}_4} \frac{1}{z^{}_2-z^{}_3}
\right] \times  \nonumber \\
& & (z^{}_1-z^{}_2)^2(z^{}_1-z^{}_3)^2(z^{}_1-z^{}_4)^2(z^{}_2-z^{}_3)^2(z^{}_2-z^{}_4)^2(z^{}_3
-z^{}_4)^2 \exp\left(-\sum^{}_i\frac{z_i^2}{4\ell_0^2}\right).
\label{fourParticle}
\end{eqnarray}
One can see that the highest power of any $z^{}_i$, e.g., $z^{}_1$, in the above expression is 5. 
In general, for the number of electrons $N^{}_e$, the highest power is $2N^{}_e-3$. In the 
thermodynamic limit, it gives the required filling factor $1/2$. 

The Pfaffian state has a very unique property in terms of the charged excitations. In fact, the 
function that describes creation of two positively charged holes has the following form 
\begin{equation}
\Psi^{}_{\mbox{Pf}}=\mbox{Pf}\left[\frac{(z^{}_i-\eta^{}_1)(z^{}_j-\eta^{}_2)+ (z^{}_j-
\eta^{}_1)(z^{}_i-\eta^{}_2)}{z^{}_i-z^{}_j}\right]         
\prod^{}_{i<j} (z^{}_i-z^{}_j)^2 \exp\left(-\sum^{}_i\frac{z_i^2}{4\ell_0^2}\right).
\label{Pf_12_holes}
\end{equation}
Here $\eta^{}_1$ and $\eta^{}_2$ are the coordinates of the holes. The charge of these excitations 
is fractional, $e^*=e/4$, and they obey the `non-Abelian' statistics \cite{RMP_nonabelions_FQHE,
Stern_Halperin_PRL_non_abelian,Stern_review_anyons}. The non-Abelian statistics means that when two 
excitations exchange their positions, it changes not only the phase of the wave function but also 
introduces an unitary transformation of the wave function itself. In general, these unitary 
transformations do not commute and the system of these excitations is called non-Abelian. 

The charged excitations of the Pfaffian state also carry the signature of Majorana fermions (MFs)
\cite{Read_Green_PRB_majorane,Ivanov_PRL_majorane}. The Majorana fermion is a special type of `particle', 
which can be seen as its own anti-particle. If $\gamma^{}_i$ is an operator corresponding to the
Majorana fermion $i$, then it satisfies the relations 
\begin{eqnarray}
& & \left\{ \gamma^{}_i, \gamma^{}_i \right\} =2 \delta^{}_{ij} \\
& & \gamma_i^\dagger = \gamma^{}_i,
\end{eqnarray}
where $\left\{ \gamma^{}_i, \gamma^{}_i \right\}$ is the anti-commutator of $\gamma^{}_i$ and 
$\gamma^{}_j$. The first relation determines that these particles are fermions and the second 
relation tells us that the particle is its own anti-particle. The Majorana particle can be 
formally expressed as a combination of the creation, $c_i^\dagger $, and annihilation, $c^{}_i$, 
operators of regular fermions,
\begin{equation}
\gamma^{}_i = c^{}_i + c_i^\dagger. 
\end{equation} 
It means that the Majorana operator creates simultaneously a particle, e.g., an electron with the 
negative charge $-e$, and an anti-particle, e.g., a hole with positive charge $e$. In conventional 
systems such an operation is impossible, but in systems such as the superconductors which 
support Cooper pairs and for which the electric charge is conserved mod 2, the Majorana fermions 
are possible. Superposition of two Majorana fermions correspond to a fermionic state where the constituent 
MFs are spatially separated. This intrinsic non-local nature of the fermionic state makes the state
being protected from local perturbations. This property and the associated non-abelian exchange statistics makes 
the MFs well suited for fault-tolerant quantum computing 
\cite{Kitaev_quantum_computer,Majorana_quantum_computer,qubits_majorana_5_2_PRL}.

Many numerical studies of the $\nu =1/2$ state have been performed in the spherical geometry 
(see, e.g., \cite{Fano_spherical_geometry}). In this geometry, the electrons are placed on a surface of a 
sphere and the magnetic field is created by a magnetic monopole placed at the center of the 
sphere. The strength of the magnetic field is characterized by the parameter $S$, where $2S$ is the number 
of magnetic fluxes through the sphere in units of the flux quantum. At the same time the parameter $S$ 
is equal to the angular momentum of single-particle states. All single-particle states have the 
same energy and form the Landau level, where different states are distinguished by the $z$ 
component of the angular momentum. Therefore, in spherical geometry, the total number of states in the 
Landau level is $2S+1$. In a many-electron system the electron-electron interaction is introduced 
through the Haldane pseudopotentials and the number of electrons, $N^{}_e$ determines the 
filling factor of the Landau level. Here the $\nu=1/2$ FQHE corresponds to the following condition: 
$2S = 2N^{}_e -3$. Although in this case the ratio of the number of electrons and the number of 
single-particle states is not 1/2, i.e., $N^{}_e/(2S+1) = N^{}_e/(2N^{}_e -2)\neq 1/2$, for  
this value of $S$, the system has a gap and the ground state wave function is close to the Pfaffian 
function. In the thermodynamic limit, i.e., for $N\rightarrow \infty$, there is a correct identity  
$N^{}_e/(2S+1) = N^{}_e/(2N^{}_e -2)\rightarrow 1/2$.

The position of an electron in spherical geometry can be described by either the two angles, $\phi$ 
and $\theta$, or two components of a spinor, $u=\cos \theta \sin \phi $ and $v$. 
In terms of the variables $u$ and $v$, the Pfaffian function takes the following form 
\begin{equation}
\Psi^{}_{\mbox{Pf}}=\mbox{Pf}\left[\frac1{u^{}_i v^{}_j -u^{}_j v^{}_i}\right]         
\prod^{}_{i<j} (u^{}_i v^{}_j -u^{}_j v^{}_i)^2. 
\label{Pf_sphere_12}
\end{equation}
In the spherical geometry the Pfaffian function is an exact wave function of the ground 
state for the system with three-particle interaction defined as 
\begin{equation}
H^{}_{int} = \frac{e^2}{\kappa \ell^{}_0 } \sum^{}_{i<j<k} 
P^{}_{ijk}( 3S -3), 
\end{equation}
where $P^{}_{ijk}$ is the three-particle projection operator onto the state with the total angular momentum $L$.

In a planar geometry, the Pfaffian function is the exact ground state wave function of the system 
with the three-body interaction \cite{PRL_5_2_numerical_Greiter,Paired_Hall_states_Nuclear_Physics} 
given by the following expression 
\begin{equation}
H^{}_{3B} = V^{}_0 \sum_{i<j<k}^{N^{}_{el}} S^{}_{ijk} \nabla_i^2 \delta(i-j) \delta(i-k),
\label{three_particle}
\end{equation}
where $\delta(i-j) = \delta(\vec{r}^{}_i - \vec{r}^{}_j) $ is the $\delta$ function, $\nabla^{}_i$ is 
the partial derivative with respect to $\vec{r}^{}_i$, and  $S^{}_{ijk}$ denotes symmetrization 
over permutations within $(ijk)$: $S^{}_{123}=f^{}_{123}+f^{}_{231}+f^{}_{312}$, and $f$ is symmetric
in its first two indices. To understand why for the three-body interaction potential 
(\ref{three_particle}) the energy of the Pfaffian state is zero, we need to look at the structure 
of the Pfaffian wave function. For example, by looking at the four-particle system 
(see Eq.~(\ref{fourParticle})) we can see that the terms with $z^{}_1$ are of the type: 
$(z^{}_1-z^{}_2)(z^{}_1-z^{}_3)^2(z^{}_1-z^{}_4)^2\ldots $. Therefore, they have one factor 
with linear dependence on $z^{}_1$, $(z^{}_1-z^{}_2)$, and all other factors have quadratic 
dependence on $z^{}_1$, $(z^{}_1-z^{}_3)^2(z^{}_1-z^{}_4)^2\ldots $. Now let us consider the 
following term in the three-particle Hamiltonian (\ref{three_particle}): $\nabla_1^2 \delta(1-2) 
\delta(1-3)$. Then to find the contribution to the energy due to that term, we need to calculate 
the second derivative of $(z^{}_1-z^{}_2)(z^{}_1-z^{}_3)^2(z^{}_1-z^{}_4)^2\ldots $ 
with respect to $z^{}_1$ and then set $z^{}_1$ equals to $z^{}_2$ and $z^{}_3$, which finally 
gives us zero. Analyzing different terms both in the Hamiltonian (\ref{three_particle}) and the 
Pfaffian wave function, we can conclude that the energy of the Pfaffian state is  zero and it is an 
eigenfunction of the three-particle Hamiltonian (\ref{three_particle}). 

\begin{figure}
\begin{center}\includegraphics[width=0.47\textwidth]{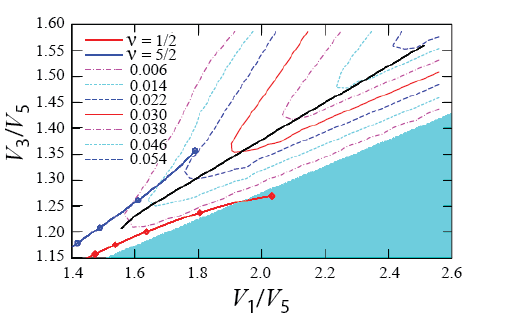}\end{center}
%\vspace*{-1cm}
\caption{ Contour plot of the FQHE gap as a function of $V^{}_1/V^{}_5$ and $V^{}_3/V^{}_5$. 
The black line marks the maximum of the overlap of the ground state wave function and the 
Pfaffian function. The lines with dots depict the effect of the finite width of a 2D layer. 
Here the red line corresponds to filling factor $\nu = 5/2$, which is the same as  $\nu=1/2$ 
at the $n=1$ Landau level, while the blue line corresponds to $\nu =1/2$ at the $n=0$ Landau 
level. The dots denote width $w/\ell^{}_0 = 0,1,2,3,4$ (from right to left). The compression 
region is shaded blue.  Reproduced with permission from 
\cite{Pfaffian52PRL104_076803}. }
\label{Fig_phase_original}
\end{figure}

For the real two-body interaction, whether the Pfaffian function is the ground state wave function 
or not is determined by the values of Haldane pseudopotentials, mainly by the values of 
$V^{}_1/V^{}_5$ and $V^{}_3/V^{}_5$. The diagram, which shows under what values of these parameters 
the half-filled Landau level becomes incompressible and is described by the Pfaffian function, 
is shown in Fig.~\ref{Fig_phase_original} \cite{Pfaffian52PRL104_076803}. The diagram depicts  
the contour plot of the gap of the system, where for the compressible state the gap is zero. The 
gap also characterizes the stability of the corresponding $1/2$-FQHE state. Here the blue region 
corresponds to the compressible state, which does not support the $1/2$-FQHE. The maximum overlap 
of the ground state wave function with the Pfaffian function is marked by a solid black line. 

The thick lines with dots correspond to the conventional electron system with parabolic energy dispersion and 
different thickness of the 2D layer. Here the thickness of the layer changes the Haldane 
pseudopotentials, resulting in the corresponding lines in the diagram. The results of 
Fig.~\ref{Fig_phase_original} clearly show that, for the conventional system, the state in the half-filled 
$n=0$ Landau level is compressible, while that for the half-filled $n=1$ Landau level is incompressible 
and the overlap of the ground state wave function with the Pfaffian function is large. The half-filled 
$n=1$ Landau level corresponds to the total filling factor of $5/2$. 

The diagram shown in Fig.~\ref{Fig_phase_original} can be also used to analyze the properties 
of half-filled Landau levels for other systems, such as graphene monolayer and graphene bilayer. 
The Haldane pseudopotentials for these systems is different from the ones for the conventional system. 

The effective electron-electron interaction potential also depends on the Landau level mixing. 
This mixing can be important for systems with small Landau level gaps, such as the ZnO quantum 
wells \cite{Pfaffian_state_ZnO_wells_2017}. The strength of mixing is determined by the ratio 
of the Coulomb interaction and the Landau level gap. When this ratio is varied, it was shown 
by numerical analysis \cite{Pfaffian_state_ZnO_wells_2017} that, in the ZnO quantum well 
systems the topological transitions can be observed for the half-filled Landau level. Further, 
in ZnO systems a stable Pfaffian state have been predicted at filling factors $5/2$ and $7/2$ 
\cite{Pfaffian_state_ZnO_wells_2017}. 

%contour plot
%of the gap as a function of v1=v5 and
%v3=v5. The black line marks the maxima
%of the overlap. The thick lines with dots
%depict the finite width trajectory, the
%upper (lower) refers to  ¼ 5
%2 ð1
%2Þ, the
%dots denoting width w=‘0 ¼ 0, 1, 2, 3,
%4 (from right to left). The compressible
%region is shaded blue (all data for Nel ¼
%16, N ¼ 29).
%%%%%%%%%%

\section{Monolayer and  bilayer graphene in a strong magnetic field}

\subsection{Graphene monolayer}

\begin{figure}
\begin{center}\includegraphics[width=0.47\textwidth]{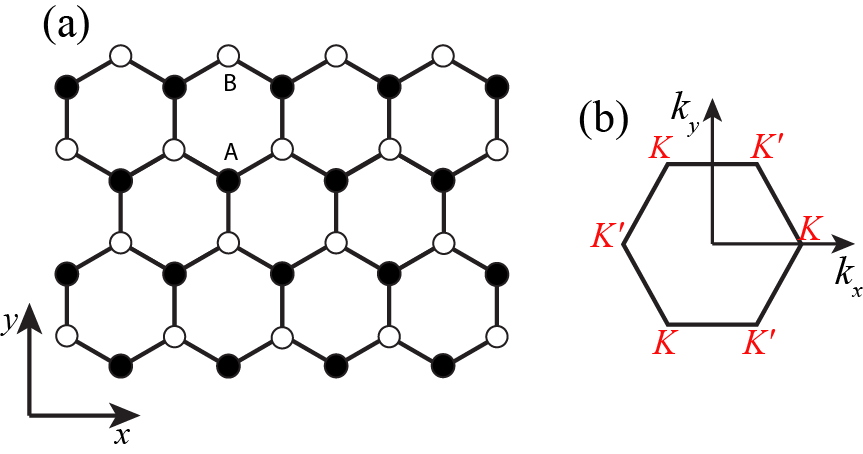}\end{center}
%\vspace*{-1cm}
\caption{ (a) Honeycomb crystal structure of graphene monolayer. Two equivalent sublattices A and B 
are marked by open and filled dots. (b) The first Brillouin zone of monolayer graphene. Two valley 
$K$ and $K^\prime$ are also shown.   }
\label{graphene_Lattice}
\end{figure}

The graphene is a monolayer of carbon atoms \cite{Graphene_Wallace_PR_1947,graphene_novoselov_2004}, 
which form the 2D honeycomb crystal structure with two sublatices, say A and B [see 
Fig.~\ref{graphene_Lattice}(a)] \cite{graphene_advances_2010}. The corresponding electron band 
structure has two valleys at two inequivalent points, $K=(2\pi/a)(\frac13,\frac1{\sqrt3})$ and 
$K^{\prime}=(2\pi/a)(\frac23,0)$, in the reciprocal space [see Fig. \ref{graphene_Lattice}(b)]. Here  
$a=0.246$ nm is the lattice constant.  The unique property of graphene is that the low-energy 
dispersion at each valley  is determined by the massless Hamiltonian of the Dirac type
(Dirac fermions) \cite{electronic_properties_graphene_RMP_2009,graphene_novoselov_2007,
graphene_advances_2010}
\begin{equation}
{\cal H}^{}_{\xi} = \xi v^{}_{\rm F}\left( 
\begin{array}{cc}
    0 & p^{}_{-} \\
    p^{}_{+} & 0  
 \end{array} 
\right),
\label{H}
\end{equation}
where $\xi$ is the valley index, which is 1 at the $K$ valley and -1 at the $K^{\prime}$ valley, 
$p^{}_- = p^{}_x- {\rm i} p^{}_y$, $p^{}_+ = p^{}_x + {\rm i} p^{}_y$,  
and  $v^{}_{\rm F} \approx 10^6$ m/s is the Fermi velocity. The corresponding eigenfunctions 
of the Hamiltonian (\ref{H}) have two components due to two sublattices, $A$ and $B$, of the graphene 
honeycomb lattice. The wave functions are expressed as $(\psi^{}_{A}, \psi^{}_B)^{T} $ for valley $K$ 
and $(\psi^{}_B, \psi^{}_A)^{T}$ for valley $K^{\prime}$, where $\psi^{}_A$ 
and $\psi^{}_B$ correspond to sublattices $A$ and $B$, respectively. 

The Landau level of electrons in graphene can be 
found from Hamiltonian (\ref{H})  by replacing 
the electron momentum $\vec{p}$ with the generalized momentum $\vec{\pi} 
= \vec{p} + e\vec{A}/c$,
\begin{equation}
{\cal H}^{}_{\xi } = \xi v^{}_{\rm F} \left( 
\begin{array}{cc}
    0 & \pi^{}_{-}  \\
    \pi^{}_{+} & 0  
 \end{array} 
\right).
\label{Hm}
\end{equation}
Then the eigenfunctions of the Hamiltonian (\ref{Hm}) have the following form  
\begin{equation}
\Psi^{K}_{n,m} = \left(\begin{array}{c}
 \psi^{}_A \\
   \psi^{}_B
\end{array}  
 \right) = C^{}_n
\left( \begin{array}{c}
 {\rm sgn}(n) {\rm i}^{|n|-1}\phi^{}_{|n|-1,m} \\
    {\rm i}^{|n|} \phi^{}_{|n|,m}   
\end{array}  
 \right),
\label{f1}
\end{equation}
for the valley $K$ ($\xi = 1$) and 
\begin{equation}
\Psi^{K^{\prime}}_{n,m} =\left(\begin{array}{c}
 \psi^{}_B \\
   \psi^{}_A
\end{array}  
 \right)=C^{}_n
\left( \begin{array}{c}
 {\rm sgn}(n) {\rm i}^{|n|-1} \phi^{}_{|n|-1,m} \\
    {\rm i}^{|n|} \phi^{}_{|n|,m}   
\end{array}  
 \right),
\label{f2}
\end{equation}
for the valley $K^{\prime}$ ($\xi=-1$). Here $C^{}_n=1 $ for $n=0$ and 
$C^{}_n=1/\sqrt2$ for $n\neq 0$ and 
\begin{equation}
\mbox{sgn}(n) = \left\{ \begin{array}{cc}
0 &  (n=0)  \\
1 &  (n>0)  \\
-1 &  (n<0).
\end{array} \right. ,
\end{equation} 
where the positive and negative values of $n$ correspond to the conduction and valence bands, respectively.
 The corresponding Landau energy spectrum takes the form \cite{Landau_levels_graphene_PRB_1956,
 electronic_properties_graphene_RMP_2009,
 graphene_in_magnetic_field_RMP_2011}
\begin{equation}
\varepsilon^{}_n=\hbar\omega^{}_B\mbox{sgn}(n)\sqrt{|n|},
\label{landau}
\end{equation}
where $\omega^{}_B=\sqrt2 v^{}_{\rm F}/\ell^{}_0$ and $\ell^{}_0=\sqrt{\hbar/e B}$ 
is the magnetic length. 

Summarizing, we can say that there are two types of Landau levels in graphene. The first one 
corresponds to $n=0$. In this case the wave function is 
\begin{equation}
\Psi^{K}_{n=0,m} =  
\left( \begin{array}{c}
 0 \\
     \phi^{}_{0,m}   
\end{array}  
 \right).
\label{f10}
\end{equation}
This wave function consists of only the $\phi^{}_{0,m}$ Landau function of a conventional system. 
The interacting electron system at $n=0$ graphene Landau level is therefore identical to the 
interacting electron system in $n=0$ Landau level of the conventional system. 

The second class of graphene Landau levels correspond to $n\neq 0$. In this case the wave function is 
\begin{equation}
\Psi^{K}_{n,m}  = \frac{1}{\sqrt{2}}
\left( \begin{array}{c}
 {\rm sgn}(n) {\rm i}^{|n|-1}\phi^{}_{|n|-1,m} \\
    {\rm i}^{|n|} \phi^{}_{|n|,m}   
\end{array}  
 \right).
\label{f1GN}
\end{equation}
Now, the graphene wave function is the mixture of $n$ and $n-1$ Landau wave functions of 
a conventional electron system. 

\begin{figure}
\begin{center}\includegraphics[width=0.47\textwidth]{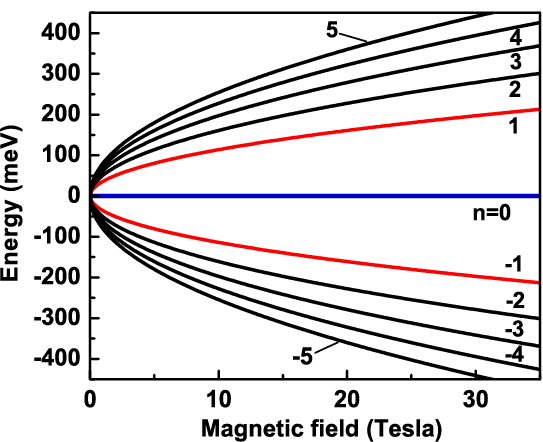}\end{center}
%\vspace*{-1cm}
\caption{ The first few Landau levels of graphene monolyaer as a function of perpendicular 
magnetic field. The Landau levels are labeled with integer index $n$, where positive and 
negative $n$ correspond to the conduction and valence bands, respectively. The red and blue 
lines mark the Landau levels for which the FQHE can be observed.   }
\label{FigLLs}
\end{figure}

For Dirac fermions in graphene, the unique feature of the Landau levels (\ref{landau}) is their 
square-root dependence on both the
magnetic field $B$ and the Landau level index $n$. This behavior is different 
from that in conventional semiconductor 2D systems with 
the parabolic energy dispersion discussed above, where the Landau levels have linear 
dependence on both the magnetic field and the Landau level index [see Eq.~(\ref{LConv})]. The 
Landau level of graphene are shown in Fig.~\ref{FigLLs} as a function of the magnetic field. 
Here the positive and negative Landau level indices $n$ correspond to the conduction and valence 
bands, respectively. These unique properties of the Landau levels of Dirac fermions in graphene 
result in the unconventional Quantum Hall Effect in graphene with quantum Hall plateaus at filling 
factors $4\left(n+\frac12\right)$ \cite{graphene_novoselov_2004,QHE_graphene_nature_2005}. 

\subsection{Bilayer graphene}

\begin{figure}
\begin{center}\includegraphics[width=0.47\textwidth]{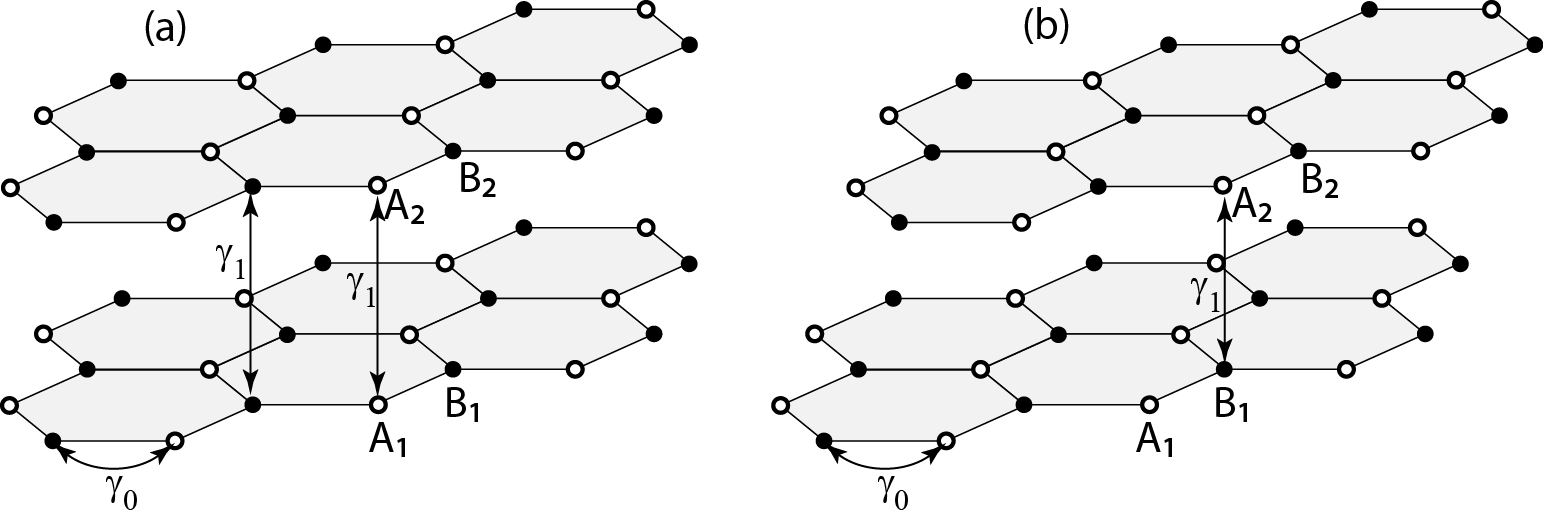}\end{center}
%\vspace*{-1cm}
\caption{ (a) Bilayer graphene with AA stacking. Here the interlayer hopping couples sublattice 
A (B) in the top layer with sublattice A (B) in the bottom layer. 
(b) Bilayer graphene with AB staking.  For such staking, sublattice A (B) in the top layer is 
coupled to sublattice B (A) in the bottom layer. }
\label{bilayer}
\end{figure}

The bilayer graphene consists of two coupled graphene monolayers 
 \cite{Falko_Landau_levels_bilayer_graphene,bilayer_graphene_review_2012}, which can be in two 
possible main stackings: (i) AA stacking and (ii) Bernal (AB) stacking, which are shown schematically in 
Fig.~\ref{bilayer}. 

For bilayer graphene with AA stacking, there is a interlayer coupling between the 
Landau levels of two layers with the same Landau level indices. Such coupling changes the energies of the 
Landau levels of graphene monolayers, but does not affect the wavefunctions of the layers. Therefore, 
the bilayer graphene Haldane pseudopotentials, which 
characterize the electron-electron interaction properties, are completely identical to the corresponding 
pseudoptentials of monolayer graphene. 

For bilayer graphene with Bernal (AB) stacking, the interlayer coupling strongly modifies the 
properties of the Landau level in the system. The corresponding Hamiltonian for valley $\xi =\pm 1$  
has the form \cite{Falko_Landau_levels_bilayer_graphene} 
\begin{equation}
{\cal H}_{\xi}^{(AB)} = \xi\left( 
\begin{array}{cccc}
\frac{U}2  & v^{}_{\rm F} \pi^{}_{-} & 0 & 0 \\
 v^{}_{\rm F} \pi^{}_{+} & \frac{U}2 &
\xi\gamma^{}_1 & 0 \\ 
 0 &\xi\gamma^{}_1 & -\frac{U}2 &
v^{}_{\rm F}
\pi^{}_{-} \\      
 0 & 0 & v^{}_{\rm F} \pi^{}_{+} & -\frac{U}2 
 \end{array} 
\right).
\label{HAB2}
\end{equation}
where $U$ is the inter-layer bias voltage and $\gamma^{}_1 \approx 400$ meV is the interlayer 
hopping integral. The corresponding wave functions have the structure of $(\psi^{}_{A_1}, 
\psi^{}_{B_1}, \psi^{}_{A_2}, \psi^{}_{B_2})^{T} $, where $A^{}_1$, $B^{}_1$ correspond to the 
lower monolayer and $A^{}_2$, $B^{}_2$ correspond to the upper monolayer.
From Hamiltonian (\ref{HAB2}), the Landau level wave functions are
\begin{equation}
\Psi^{\rm (bi)}_{n,m}  = 
\left( \begin{array}{c}
 \xi  C^{}_1 \phi^{}_{n-1,m} \\
   C^{}_2   \phi ^{}_{n,m} \\  
   C^{}_3  \phi^{}_{n,m} \\
  \xi C^{}_4 \phi ^{}_{n+1,m}  
\end{array}  
 \right),
\label{fAB2}
\end{equation} 
where $C^{}_1$, $C^{}_2$, $C^{}_3$, and $C^{}_4$ are constants. The wave functions in 
the bilayer graphene with AB stacking are therefore the mixtures of the conventional Landau 
wavefunctions with indices $n-1$, $n$, and $n+1$. 

In Eq.~(\ref{fAB2}) the Landau index $n$ can take the following values: $n=-1,0,1,\ldots$. Here 
we assume that if the index of the Landau wave function, $\phi ^{}_{n,m}$, is negative 
then the function is identically zero, which means that  $\phi^{}_{-2,m} \equiv 0$ and 
$\phi^{}_{-1,m} \equiv 0$. Hence, for $n=-1$, the wave function (\ref{fAB2}) is just 
$\Psi^{\rm (bi)}_{-1,m} = (0,0,0,\phi ^{}_{0,m})$, i.e., the coefficients $C^{}_1$, $C^{}_2$, 
$C^{}_3$ are zero. Therefore, there is only one energy level corresponding to $n=-1$. 
For $n=0$, the wave function (\ref{fAB2}) has zero coefficient $C^{}_1$ and, correspondingly, 
there are only three energy levels.  

For $n> 0$, there are four eigenvalues of the Hamiltonian (\ref{HAB2}), corresponding to four Landau 
levels in a bilayer graphene at a given valley $\xi = \pm 1$. The eigenvalue equation, which 
determines the Landau levels is written as \cite{Pereira_Landau_levels_bilayer_graphene_PRB} 
\begin{equation}
\left[\left(\varepsilon+\xi u \right)^2-2n\right]
\left[\left(\varepsilon-\xi u \right)^2-2(n +1)\right] = 
\tilde{\gamma}_1^2 \left[\varepsilon^2-u^2\right],
\label{eigen1}
\end{equation}
where $\varepsilon$ is the energy of the Landau level in units of $\epsilon^{}_B$, 
$\epsilon^{}_B=\hbar v^{}_{\rm F}/\ell^{}_0$,  
$\tilde{\gamma}^{}_1=\gamma^{}_1/\epsilon^{}_B$, and $u=U/\epsilon^{}_B$. 

\begin{figure}
\begin{center}\includegraphics[width=0.47\textwidth]{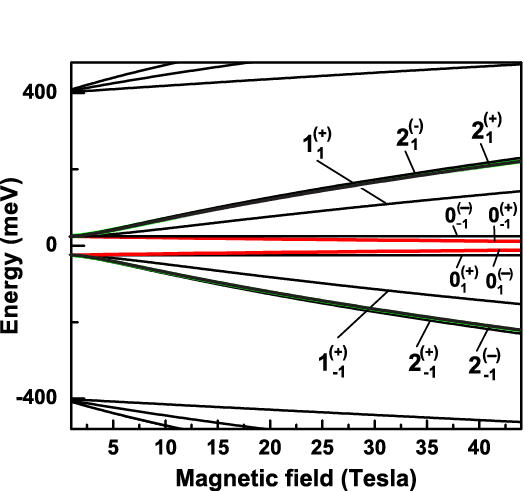}\end{center}
%\vspace*{-1cm}
\caption{ A few lowest Landau levels of bilayer graphene as a function of the magnetic field. 
The bilayer graphne has the 
AB staking. The Landau levels of both $K$ and $K^\prime$ are shown. The red lines show 
special Landau levels, at which the $\nu =1/2$ Pfaffian incompressible state can be realized. 
The bias voltage is 5 meV and interlayer hopping integral is $\gamma^{}_1 = 400 $ meV.      }
\label{Fig_energy_bilayer}
\end{figure}

The four Landau levels determined by Eq.~(\ref{eigen1}) have two levels with negative energy 
(valence band) and two levels with positive energy (conduction band).
It is then convenient to label these levels as follows: for a given value of $n$  and a given valley $\xi$ 
we label the levels as $n^{(\xi)}_i$, where $i = -2, -1, 1, 2$ in the ascending order and the  
negative and positive values of $i$ correspond to the valence and conduction bands, respectively. 
Although for $n=0$ there are only three Landau levels and for $n=-1$ there is only one Landau level, 
it is convenient to combine them in a single set of $n=0$ Landau levels and label them as 
$0^{(\xi)}_i$, where $i=-2,-1,1,2$. Also, the Landau levels of different valleys are related as follows 
$\epsilon (n^{(\xi)}_i) = - \epsilon (n^{(-\xi)}_{-i})$. 

With the known wave functions (\ref{fAB2}) of the bilayer graphene Landau levels, the form factor 
in Eq.~(\ref{Vm}) for Haldane pseudopotentials can be obtained from the following expression 
\begin{equation}
F^{}_n(q) = |C^{}_1|^2 L^{}_{n-1}(q^2/2) +  
\left(|C^{}_2|^2 + |C^{}_3|^2\right) L^{}_{n}(q^2/2) 
+ |C^{}_4|^2 L^{}_{n+1}(q^2/2).
\label{FFbilayer}
\end{equation}

%First, we consider the case of zero bias voltage. The fours Landau levels are given by the following expression 
%\begin{equation}
%\epsilon = \pm \sqrt{ 2n+1 + \frac{\tilde{\gamma}^{2}_1 }{2} \pm 
%\frac12 \sqrt{(2+ \tilde{\gamma}^{2}_1 )^2 + 8 n \tilde{\gamma}^{2}_1  }   },
%\label{zero}
%\end{equation}
%where each level has two-fold valley and two-fold 
%spin degeneracy. 

%Since the FQHE is expected only in the Landau levels with low values of the index, $n$, we 
%consider below the sets of the Landau levels of bilayer graphene  with $n=0$ and $n=1$ only. 
%The wavefunctions of these Landau levels are mixtures of the conventional 
%Landau functions with indices $0$, $1$, and $2$. 

There are two special Landau levels of bilayer graphene with almost zero energy \cite{bilayer_PRL}. 
The first one corresponds to $n=-1$ with the energy of $\varepsilon = - \xi u$ and the corresponding 
wave function, 
\begin{equation}
\Psi^{\rm (bi)}_{0^{(+)}_1,m}=
\Psi^{\rm (bi)}_{0^{(-)}_{-1},m} =
\left( \begin{array}{c}
0 \\
 0 \\  
  0 \\
   \phi^{}_{0,m}  
\end{array}  
 \right).
\label{f00}
\end{equation} 
The wave function consists of only the $n=0$ conventional Landau level wave function. Therefore, 
the FQHE in this bilayer Landau level is exactly the same as the one in the $0$-th conventional 
Landau level. For example, we can say that there is no FQHE in the half-filled $0^{(+)}_1$ Landau level. 

In order to find the half-filled level with possible Pfaffian function as the ground state, the 
most interesting bilayer Landau level is $0^{(+)}_{-1}$ in the $K$ valley and 
$0^{(-)}_1 $ in the $K^\prime $ valley \cite{bilayer_PRL}. 
For small values of $U$, the energy of this state is  almost zero, $\varepsilon\approx 0$. 
The wave function of $0^{(+)}_{-1}$  Landau level has the form 
\begin{equation}
\Psi^{\rm (bi)}_{0^{}_{-1},m}  = 
 \frac1{\sqrt{\tilde{\gamma}_1^2 + 2}}
\left( \begin{array}{c}
 0 \\
 \sqrt{2} \phi^{}_{0,m} \\  
 0 \\
  \tilde{\gamma}^{}_1 \phi^{}_{1,m}  
\end{array}  
 \right) = \frac1{\sqrt{\gamma_1^2 + 2 \epsilon_B^2}}
\left( \begin{array}{c}
 0 \\
 \sqrt2 \epsilon^{}_B\phi^{}_{0,m} \\  
 0 \\
  \gamma^{}_1 \phi^{}_{1,m}  
\end{array}  
 \right).
\label{f11}
\end{equation} 
The wave function is the mixture of $\phi^{}_{0,m}$ and 
$\phi^{}_{1,m}$ states.  For a relatively small magnetic field, $\epsilon^{}_B \ll \gamma^{}_1$, the 
wave function becomes $( 0, 0, 0, \psi^{}_{1,m})^T$, i.e., it is completely identical to the  $n=1$ 
conventional Landau level. In a large magnetic field, $\epsilon^{}_B \gg \gamma^{}_1$, the $0^{(+)}_{-1}$ 
Landau  wave function becomes $(0, 0, \psi^{}_{0,m}, 0)^T$, i.e., it is the same as $n=0$ conventional
Landau level. The interesting aspect of this is that, by varying the magnetic field the electron-electron 
interactions within the $0^{(+)}_{-1}$ Landau level transform from $n=1$ to the conventional $n=0$ case. 

\section{Pfaffian states in graphene}

\subsection{Graphene monolayer}

The electron-electron interactions and correspondingly the nature of the ground state within a 
given Landau level are determined by the Haldane pseudopotentials, which are defined by the 
corresponding form factor $F(q)$. For graphene monolayer there are two types of Landau levels: 
(i) the $n=0$ Landau level and (ii) the $n\neq 0$ levels. For the $n=0$ Landau level, the form factor
is given by
\begin{equation}
F^{}_0(q) =  L^{}_{0}(q^2/2)=1.
\label{FGM1}
\end{equation}
This is exactly the same form factor as the one for the $n=0$ conventional Landau level, for which 
there is no $\nu=1/2$ FQHE state. We can then conclude that in the $n=0$ graphene Landau level, 
there is no incompressible $\nu=1/2$-FQHE state with the Pfaffian function as a ground state wave 
function. At the same time, all other FHQE states, such as $p/q$ with odd $q$, have exactly the 
same properties, i.e., the same ground state wave functions and the same excitation gaps, as the 
ones for the $n=0$ conventional Landau level. This suggests that, as long as there is no inter-Landau 
level mixing, the electron systems in the $n=0$ graphene Landau level are identical to the systems 
in the $n=0$ conventional Landau level. 

On the other hand, for the $n\neq 0$ graphene Landau level, the form factor is 
\begin{equation}
F^{}_n(q) =  \frac{1}{2} \left( L^{}_{n-1}(q^2/2) +   L^{}_{n}(q^2/2) \right) .
\label{GM2}
\end{equation}
For example, for $n=1$, $L^{}_0(q^2/2)=1$ and  $L^{}_0(q^2/2)=1 -q^2/2$ and the form factor is 
$F^{}_n(q) = 1- q^2/4$. In this case the form factor is the mixture of the form factors of $n$ 
and $n-1$ conventional Landau levels. The electron-electron interactions in this case are 
completely different from those in conventional systems. One of these differences is related to 
stability, i.e., the magnitude of the excitation gaps of the conventional FQHE with the 
filling factor $p/q$, where $ q$ is odd. This means that, out of all the graphene Landau levels, 
including the $n=0$ Landau level, the largest FQHE gaps are realized in the $n=1$ Landau level 
\cite{Apalkov_FQHE_graphene_PRL,FQHE_in_graphene_aspects_review}. This is different from 
conventional systems, where the conventional FQHE states have the largest gaps for the 
$n=0$ Landau level.  

Our extensive numerical analyses have established that there is no incompressible $\nu =1/2$ 
state of the Pfaffian type for $n= 0$ and $n=1$ graphene Landau levels \cite{Apalkov_FQHE_graphene_PRL}. 
Here, the excitation gap is small and the overlap of the ground state wave function with the Pfaffian 
function is also small. This analysis has been done without considering inter-Landau level mixing. 
This mixing can alter the properties of the half-filled Landau level, resulting in the formation of 
an incompressible state. For example, for conventional systems it was shown that the inter-Landau mixing 
favors the anti-Pfaffian state over other type of ground states of the half-filled $n=1$ Landau level
\cite{LL_mixing_5_2_PRL}. Also, both for conventional and graphene systems, the Landau level mixing 
crucially modifies for the $1/2$-FQHE state the Haldane pseudoptentials $V_1$, $V_3$, and $V_5$
\cite{realistic_Hamiltonian_5_2}. The Landau level mixing is also more pronounced at higher Landau levels. 
It should be noted that, recent experimental results \cite{graphene_1_2_high_Landau_exp_2018} indeed suggest 
that the incompressible $\nu=1/2$ state may be realized in monolayer graphene, albeit in higher $n=3$ 
Landau level. Here, the inter-Landau mixing can be important and it can be the reason for the incompressible 
$\nu=1/2$ state being present in the higher Landau level in graphene. Further, the even-denominator 
incompressible states have been observed experimentally in low Landau levels in graphene within some 
range of the magnetic fields \cite{even_denominator_graphene_high_LL_exp_2017}, but the origin of those 
states is yet to be clearly determined and it can perhaps be related to multicomponent fractional 
quantum Hall states.

\subsection{Bilayer graphene}

For the general bilayer Landau level, the form factor is given by the expression (\ref{FB01}) below. 
Due to the bilayer nature of the system, there are extra parameters which can control the 
form factor and correspondingly the interaction strength. These parameters are the bias voltage 
and the direction of the magnetic field, i.e., the component of the magnetic field that is parallel 
to the bilayer. As we have mentioned above, in bilayer graphene there are two 'special' Landau 
levels $0_{-1}^{(+)}$ ($K$ valley) and $0_1^{(-)}$ ($K^\prime$ valley) for which the Pfaffian 
function can be the ground state of the half-filled Landau level 
\cite{bilayer_PRL,Dirac_fermions_monolayer_bilayer_graphene_Apalkov_Chakraborty}. 

 \begin{figure}
\begin{center}\includegraphics[width=0.47\textwidth]{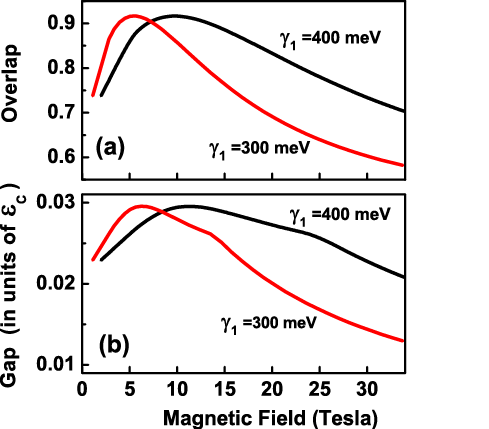}\end{center}
%\vspace*{-1cm}
\caption{ 
Panel (a): Overlap of the ground state wave function of $\nu = 1/2$ state with the Pfaffian 
state as a function of the magnetic field. Panel (b): the collective excitation gap of the $\nu= 1/2$ 
state as a function of the magnetic field.   
The data are shown for the Landau level $0_{-1}^{(+)}$ of the bilayer graphene and for two 
values of the interlayer hopping integral, $\gamma^{}_1$.  Reproduced with permission from 
\cite{bilayer_PRL}.
}
\label{FigPf2_overlap_gap}
\end{figure}

Extensive studies of the finite-size system of bilayer graphene have been performed by us in 
the spherical geometry \cite{Haldane_PRL_1983,Haldane_Rezayi_PRL_1985,Fano_spherical_geometry}. 
Those studies have revealed that for all bilayer Landau levels, except 
the ones $0_{-1}^{(+)}$ and $0_1^{(-)}$, the overlap of the $\nu=\frac12$ ground state with the 
Pfaffian state is relatively small (less than $0.5$). Therefore, for those Landau levels, 
the $\nu=\frac12$ system is gapless and compressible. A different situation occurs for the Landau 
levels $0_{-1}^{(+)}$ and $0_1^{(-)}$. These levels belong to the $K$ and $K^\prime$ valleys, 
respectively, and have exactly the same interaction properties. Hence, it is enough to consider 
only one of these levels, e.g.,  $0_{-1}^{(+)}$, as for the other Landau level, $0_1^{(-)}$, the 
results are exactly the same. 

For the $0_{-1}^{(+)}$ Landau level the form factor is 
\begin{equation}
F^{}_{0^{}_{-1}}(q)=\left[\frac{\gamma_1^2}{\gamma_1^2+2\epsilon_B^2}\right] 
L^{}_1(q^2/2) + \left[\frac{2\epsilon_B^2}{\gamma_1^2+2\epsilon_B^2 }
\right] L^{}_0(q^2/2).
\label{FB01}
\end{equation} 
The strength of magnetic field is determined by the parameter $\epsilon^{}_B$. With increasing 
magnetic field, i.e., with increasing $\epsilon^{}_B$, the form factor of the $0_{-1}^{(+)}$ level 
goes through the following main regions: (i) for small $B$, $\epsilon^{}_B \ll \gamma^{}_1$, the 
form factor is $L^{}_1(q^2/2)$ and identical to the one of the $n=1$ conventional Landau level; 
(ii) at $\epsilon^{}_B=\gamma^{}_1/\sqrt{2}$, the form factor is $\frac12 
[L^{}_0(q^2/2) + L^{}_1(q^2/2)]$ and is the same as the one of the $n=1$ Landau level of monolayer 
graphene; (iii) at $\epsilon^{}_B \gg \gamma^{}_1$, the form factor is 
 $L^{}_0(q^2/2)$ and is the same as the one of the $n=0$ conventional Landau level.

Since the $\nu=1/2$ incompressible state is realized only in the $n=1$ conventional Landau level, 
then for bilayer graphene, the $\nu=1/2$ Pfaffian state should be observed only for relatively small 
values of the magnetic field. At the same time, it so happens that as a function of the magnetic 
field, the characteristics of the $\nu = 1/2$ system is not monotonic. In Fig.~\ref{FigPf2_overlap_gap}
(a) the overlap of the $\nu =1/2$ wave function with the Pfaffian function is shown as a function 
of the magnetic field. The calculations were done in a spherical geometry for $N^{}_e  = 14$ 
electron system. The overlap has a maximum at a finite magnetic field. The position of the maximum 
depends on the interlayer hopping integral, $\gamma$. For example, for $\gamma=400 $ meV, the 
maximum overlap occurs for $\approx 10 $ T. The profile of the overlap is also correlated with 
the gap of the collective excitation of the system [see Fig.~\ref{FigPf2_overlap_gap}( b)]. 
The energy gap also has a maximum at a finite magnetic field. Since in a small magnetic field, 
$B\rightarrow 0$, the $0_{-1}^{(+)}$ bilayer Landau system is equivalent to $n=1$ conventional
Landau systems, we are allowed to conclude that the stability of the  Pfaffian $\nu =\frac12$ state 
in bilayer graphene can be increased compared to that of the conventional systems. 

The position of the maximum in Fig.~\ref{FigPf2_overlap_gap} can be analyzed further by looking 
at some dimensionless quantities. For instance, in dimensionless units the maximum is achieved 
at $\gamma_1/\epsilon_B \approx 4.9$. This means that, in terms of the magnetic field, the position 
of the maximum of the overlap is approximately proportional to $\gamma_1$, which is also seen in 
Fig.~\ref{FigPf2_overlap_gap}. 

\begin{figure}
\begin{center}\includegraphics[width=0.47\textwidth]{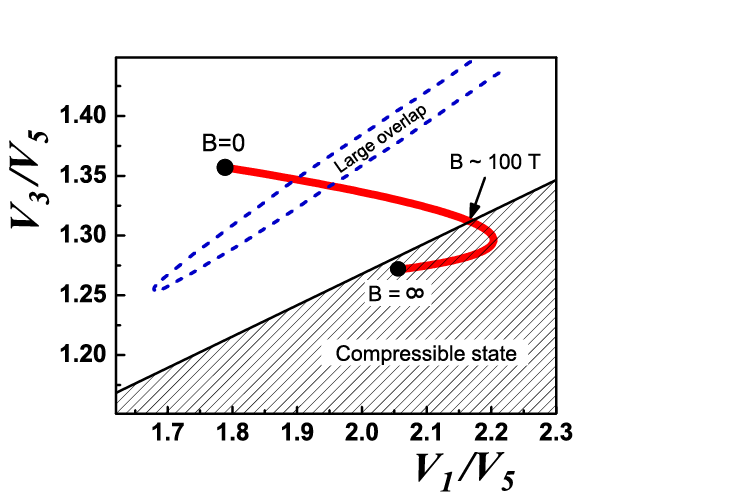}\end{center}
%\vspace*{-1cm}
\caption{ 
Trajectory of the inter-electron interaction with varying magnetic field. 
The trajectory is shown by a solid red line in the plane $(V^{}_1/V^{}_5)-(V^{}_3/V^{}_5)$ 
for the Landau level $0_{-1}^{(+)}$ of the bilayer graphene.
The initial point of the trajectory (at $B = 0$) corresponds
to the conventional system at $n = 1$ Landau level, 
while the final point (at $B = \infty $) corresponds to the conventional system at the 
$n = 0$ LL. The hatched region illustrates the compressible $\nu = 1/2$ state, while the 
blank region corresponds to the incompressible $\nu = 1/2$ state following the 
Ref.~\cite{Pfaffian52PRL104_076803} and Fig.~\ref{Fig_phase_original}. The crossing of the
boundary between the compressible and incompressible states
occurs at $B \sim 100$ Tesla for 
the interlayer hopping integral $\gamma^{}_1 = 400$ meV.
The blue dashed line shows the region of large overlap of the ground state wave function with
the Pfaffian function. Reproduced with permission from 
\cite{bilayer_PRL}.
}
\label{Fig4_phase}
\end{figure}

Using the phase diagram shown in Fig.~\ref{Fig_phase_original}, which characterizes the stability 
of the $\nu = 1/2$ state described by the Pfaffian function, we can illustrate how the interaction 
parameters in the $0_{-1}^{(+)}$ Landau level change with the magnetic field [see Fig.~\ref{Fig4_phase}]. 
For a small magnetic field, the ground state of the system is well described by the Pfaffian function, 
while the overlap and the corresponding excitation gap reach their maximum at intermediate values 
of the magnetic field and finally, for a large magnetic field, $> 100$ T, the electron system 
becomes compressible.  

Another parameter that can be used to control the stability of the Pfaffian ground state, is 
the magnetic field that is {\it parallel} to the 2D layer 
\cite{Dirac_fermions_monolayer_bilayer_graphene_Apalkov_Chakraborty}. The reason why the parallel 
magnetic field changes the wave functions and the interaction potential is due to the bilayer nature of 
the system. In this case the system has extra dynamics in the $z$ direction, which is described 
in the model as an inter-layer hopping. The parallel component of the magnetic field can be introduced 
into the Hamiltonian of bilayer graphene through a Peierls substitution as an extra 
position-dependent phase factor in the inter-layer hopping integral. The details of that study
can be found elsewhere \cite{Dirac_fermions_monolayer_bilayer_graphene_Apalkov_Chakraborty}.

\begin{figure}
\begin{center}\includegraphics[width=0.47\textwidth]{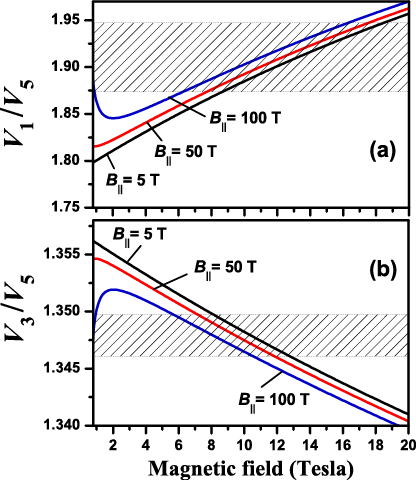}\end{center}
%\vspace*{-1cm}
\caption{  Ratios of pseudopotentials at two values of angular momentum 
$V_1/V_5$ [panel (a)] and $V_3/V_5$ [panel (b)]  are 
shown as the function of perpendicular component of magnetic field and 
for different parallel components of magnetic field, $B_{\parallel} = 5$ T, 50 T, and 
100 T. The data are shown for Landau level $0_{-1}^{(+)}$ of bilayer graphene. 
The hatched regions correspond to the values of pseudopotentials for which 
the large overlap of the ground state wave function with 
the Pfaffian function and large excitation gap of incompressible ground state are achieved. 
Reproduced with permission from \cite{Dirac_fermions_monolayer_bilayer_graphene_Apalkov_Chakraborty}.  }
\label{Fig_ratio}
\end{figure}

The effect of the parallel component of the magnetic field is illustrated in Fig.~\ref{Fig_ratio}, 
where the parameters $V_1/V_5$ and $V_3/V_5$  are shown as a function of the perpendicular component 
of the magnetic field and for different values of its parallel component. The most stable Pfaffian 
state is illustrated by the dashed lines. The results demonstrate that with increasing parallel 
component of the magnetic field the values of pseudopotentials, which correspond to the most stable 
Pfaffian state, are realized for smaller values of the perpendicular component of the magnetic field. 
Therefore, the parallel component of the magnetic field can be used to tune the stability of the 
Pfaffian state and the magnetic field where it can be realized.  

\subsection{Finding the Pfaffians}

Even-denominator fractional quantum Hall effect has been observed experimentally in bilayer graphene 
\cite{Experimental_1_2_bilayer_Nature_2017,Experimental_1_2_bilayer_Science_2017}. It was found that 
the corresponding states were spin-polarized and it was suggested that they are Pfaffian or 
anti-Pfaffian states corresponding to the half-filled Landau level. Such states were observed at 
high energy bilayer Landau levels that are a mixture of the $n=1$ and $n=0$ conventional Landau levels. 
It was noted that the FQHE was observed only at the filling factor $\nu =3/2$, but not at $\nu=1/2$ 
and $5/2$ \cite{Experimental_1_2_bilayer_Science_2017}. This observation illustrates 
the high sensitivity of the half-filled incompressible states to the structure of the electron 
wave functions which determine the electron-electron interaction potentials. Also the measured 
activation gap of the half-filled incompressible state was several times larger than the largest 
gaps measured in the GaAs systems \cite{Experimental_1_2_bilayer_Nature_2017}, which also illustrates 
that the half-filled FHQE state is much more stable in bilayer graphene then in conventional electron 
systems. It was observed that the stability of the experimentally found $1/2$-FQHE states is sensitive 
to both the magnitude of the magnetic field and the perpendicular component of the electric field. 
With variation of these parameters, transitions from the compressible to the incompressible states have 
been observed \cite{Experimental_1_2_bilayer_Science_2017}. Such a behavior is in complete
accord with our theoretical predictions elaborated in the previous section. 

\section{Conclusion}

Two-dimensional electron systems placed in a strong magnetic field can support charged excitations 
with fractional charge and non-Abelian statistics. These excitations are possible only due to the
nature of electron-electron interactions of a special type. The example of such a system is an electron gas 
in a given Landau level that is only half filled. Under a special profile of the interaction potential, 
the ground state of the $\nu=1/2$-Landau level is described by the Pfaffian function. Here the 
profile of interaction potential is determined by the form factor of the corresponding Landau level. 
As a result, for the conventional electron systems, the incompressible $\nu =1/2$-FQHE state is 
realized only in the $n=1$ Landau level, but for the graphene monolayer which has the relativistic-like
low-energy dispersion (that of the Dirac fermions), there is no incompressible $\nu =1/2$ Pfaffian state 
in any Landau level. This theoretical conclusion is based on the properties of electrons within a 
given Landau level, without considering the admixture of other Landau levels. The Landau level mixing, 
which is especially relevant for higher Landau levels, can modify the properties of the half-filled Landau 
states [see \cite{LL_mixing_5_2_PRL,Breaking_Particle_Hole_Symmetry_5_2_PRL_2011,%
Landau_level_mixing_perturbative_limit_PRB_2013,LL_mixing_phase_diagram_5_2,%
realistic_Hamiltonian_5_2,Pfaffian_state_ZnO_wells_2017}. It can also open up the possibility to observe 
the $\nu=1/2$-FQHE in higher Landau levels in graphene \cite{graphene_1_2_high_Landau_exp_2018}. 

In this context, bilayer graphene is truly unique because for bilayer graphene with AB staking, there are 
two special Landau levels: one in the $K$ valley and another one in the $K^\prime $ valley, for which, 
there is a range of magnetic fields where the $\nu =1/2$ ground state is determined by the Pfaffian 
function. Stability of such a ground state, i.e., its collective excitation gap, depends on the 
magnetic field and for a finite magnetic field, the $\nu =1/2$-FQHE state becomes even more stable 
than the corresponding state in a conventional electron system. Another important property of bilayer 
graphene is that there are external parameters, such as the bias voltage and the in-plane magnetic 
field, that can change the inter-electron interaction strength within a given Landau level and 
correspondingly change the properties of the $\nu=1/2$ Pfaffian state. 

In addition to the unique properties of the $\nu =1/2$ Pfaffian state, the system with half filling 
of a given Landau level has another interesting feature. In particular, when the projection on a 
given Landau level is considered, i.e., only the states within this Landau level are taken into 
account with inter-landau level mixing, then the system itself has a particle-hole symmetry. It 
means that it can be described either as the $\nu = 1/2$ particle (electron) system or the 
$\nu =1/2$ hole system. The Pfaffian function, which describes the $\nu = 1/2$ particle system, is 
not invariant under the exchange of particles and holes. When the particle-hole symmetry operation 
is applied to the Pfaffian function it generates the Pfaffian conjugated function, which is known 
as the anti-Pfaffian \cite{anti_pfaffian_PRL_2007,anti_pfaffian_2_PRL_2007}. The anti-Pfaffian 
function is given by
\begin{equation}
\Psi^{}_{\mbox{aPf}}=\mbox{Pf}\left[\frac{z^{}_i-z^{}_j}
{(z_i^*-z_j^*)^2}\right]         
\prod_{i<j} (z^{}_i-z^{}_j)^2 \exp\left(-\sum_i\frac{z_i^2}{4\ell_0^2}\right).
\label{A_Pf_12}
\end{equation}
In addition to the Pfaffian and anti-Pfaffian functions, another type of function have been proposed
in the literature. It is called the PH-Pfaffian and is symmetric with respect to the particle-hole 
symmetry operator \cite{PH_Pfaffian}. The PH-Pfaffian is given by the following expression 
\begin{equation}
\Psi_{\mbox{PH-Pf}}=\mbox{Pf}\left[\frac{1}
{z_i^*-z_j^*}\right]         
\prod_{i<j} (z^{}_i-z^{}_j)^2 \exp\left(-\sum_i\frac{z_i^2}{4\ell_0^2}\right).
\label{PH_Pf_12}
\end{equation}
All these functions, Pfaffian, anti-Pfaffian, and PH-Pfaffian, are of the superconductor-type with Cooper 
pairing but having different pairing symmetries. The corresponding charged excitations are non-Abelian anyons. 
Discussion of the conditions under which different Pfaffian-type functions are to be realized and their 
properties can be found in the literature \cite{Landau_level_mixing_perturbative_limit_PRB_2013,%
Breaking_Particle_Hole_Symmetry_5_2_PRL_2011,density_matrix_renormalization_group_5_2_PRB,%
LL_mixing_5_2_PRL,LL_mixing_PRB,LL_mixing_phase_diagram_5_2,realistic_Hamiltonian_5_2,LL_mixing_5_2_Jain_2010,%
Plasma_analogy_5_2,PH_symmetry_breaking_5_2_2018,Effective_field_theory_5_2,%
Stability_particle_hole_Pfaffian_state_Haldane_2021}. 
Confirmation of the presence of Pfaffian functions in half-filled FQHE will provide another open
door to our understanding of this distinctive many-body phenomena that have enthralled us for more
than four decades. Additionally, it would be equally exciting to watch that the work of Cayley and Pfaff finally 
see the light of day through bilayer graphene.

\bibliographystyle{cas-model2-names}
%\bibliographystyle{apsrmp}
%\bibliography{references}

\end{document}